\documentclass[a4paper]{article}

\usepackage{INTERSPEECH2019}
\usepackage{amsmath,amssymb,fixmath}
\usepackage{subfig}
\usepackage{graphicx}
\usepackage{float}
\usepackage{cite}
\usepackage[dvips]{epsfig}

\PassOptionsToPackage{hyphens}{url}\usepackage{hyperref}

\DeclareMathOperator{\E}{\mathbb{E}}
\DeclareMathOperator{\bx}{\mathbold{x}}

\DeclareMathOperator{\bz}{\mathbold{z}}
\DeclareMathOperator{\bxhat}{\mathbold{\hat{x}}}

\title{Probability Density Distillation with Generative Adversarial Networks for High-Quality Parallel Waveform Generation}
\name{Ryuichi Yamamoto$^1$, Eunwoo Song$^2$ and Jae-Min Kim$^2$}
\address{
  $^1$LINE Corp., Tokyo, Japan.\\
  $^2$NAVER Corp., Seongnam, Korea}
\email{ryuichi.yamamoto@linecorp.com, \{eunwoo.song, kjm.kim\}@navercorp.com}

\begin{document}

\maketitle
\begin{abstract}
    This paper proposes an effective probability density distillation (PDD) algorithm for WaveNet-based parallel waveform generation (PWG) systems. 
    Recently proposed teacher-student frameworks in the PWG system have successfully achieved a real-time generation of speech signals. However, the difficulties optimizing the PDD criteria without auxiliary losses result in quality degradation of synthesized speech.
    To generate more natural speech signals within the teacher-student framework, we propose a novel optimization criterion based on generative adversarial networks (GANs).
    In the proposed method, the inverse autoregressive flow-based student model is incorporated as a generator in the GAN framework, and jointly optimized by the PDD mechanism with the proposed adversarial learning method.
    As this process encourages the student to model the distribution of realistic speech waveform, the perceptual quality of the synthesized speech becomes much more natural.
    Our experimental results verify that the PWG systems with the proposed method outperform both those using conventional approaches, and also autoregressive generation systems with a well-trained teacher WaveNet. 
    
\end{abstract}

\section{Introduction}
\label{intro}

    
    Generative models using WaveNet have significantly improved the quality of synthetic speech signals \cite{Oord2016WaveNetAG}.
    In this kind of system, the time domain speech signal is represented as a sequence of discrete symbols, and its distribution is autoregressively modeled by stacked convolutional neural networks. 
    By appropriately conditioning the acoustic features to the input, WaveNet has also been successfully adopted in a neural vocoder structure for statistical parametric speech synthesis systems \cite{tamamori2017speaker, hayashi2017multi, song2019excitnet, Wang2018WNComparison}, and end-to-end speech synthesis systems \cite{Arik2017DeepVR,Char2017DeepV2,Ping2017DeepV3,Shen2018NaturalTS, kwon2019effective}.


    However, compared with traditional parametric vocoders \cite{Morise2016WORLDAV,agiomyrgiannakis2015vocaine, raitio2014voice,song2017effective}, the WaveNet's inference speed is inherently slow owing to its autoregressive model structure.
    To address this problem, \textit{teacher-student framework}-based fast waveform generation methods (e.g., parallel WaveNet and ClariNet) have been proposed \cite{Oord2018ParallelWF, Ping2018ClariNetPW}.
    In this framework, a bridge defined as probability density distillation (PDD) transfers the knowledge of a well-trained autoregressive teacher WaveNet to an inverse autoregressive flow (IAF)-based student model.
    As the architecture of feedforward IAFs enables transforming a simple noise signal to a complex distribution in parallel \cite{Kingma2016ImprovingVI}, the IAF student can generate speech waveform within a real-time speed.

    Typically, conventional PDD methods employ a minimization criterion based on the Kullback-Leibler divergence (KLD) between the output distributions of the student and teacher networks \cite{Oord2018ParallelWF}.
    However, as the objective of this criterion is to guide the student model to learn the teacher's distribution, the best achievable quality of the distilled student cannot be better than that of the teacher network.
    Although combining auxiliary losses (e.g., a frame-level power loss between recorded and synthetic speech signals) to the KLD criterion helps generating more natural speech segments \cite{Ping2018ClariNetPW}, it often suffers from unexpected artifacts in the synthesis step due to the difficulties to converge the student model.

    To further improve synthetic speech quality of WaveNet-based parallel waveform generation (PWG) systems, we propose a generalized optimization criterion for training the IAF students by incorporating generative adversarial networks (GANs) \cite{goodfellow2014generative}.
    In the proposed method, a teacher WaveNet is first obtained via maximum likelihood estimation, and an IAF student is incorporated as a generator within the GAN framework.
    Finally, all the weights in the student model are jointly optimized by the PDD mechanism with an adversarial learning method.
    Because the adversarial training encourages the IAF student to learn the distribution of realistic speech waveform, the perceptual quality of synthesized speech becomes much more natural.
    Furthermore, the joint optimization with conventional distillations addresses the difficulties of feedforward GAN to model the long term dependency of the speech signal.
    Consequently, the performance of the distilled student is effectively improved.


    We investigate the effectiveness of the proposed method by conducting subjective evaluations with the PWG systems.
    The experimental results show that the proposed adversarial training method provides much better perceptual quality than conventional approaches while maintaining the equivalent generation speed; moreover, outperforms even the autoregressive teacher WaveNet.

\begin{figure*}[t]
  \centerline{\epsfig{figure=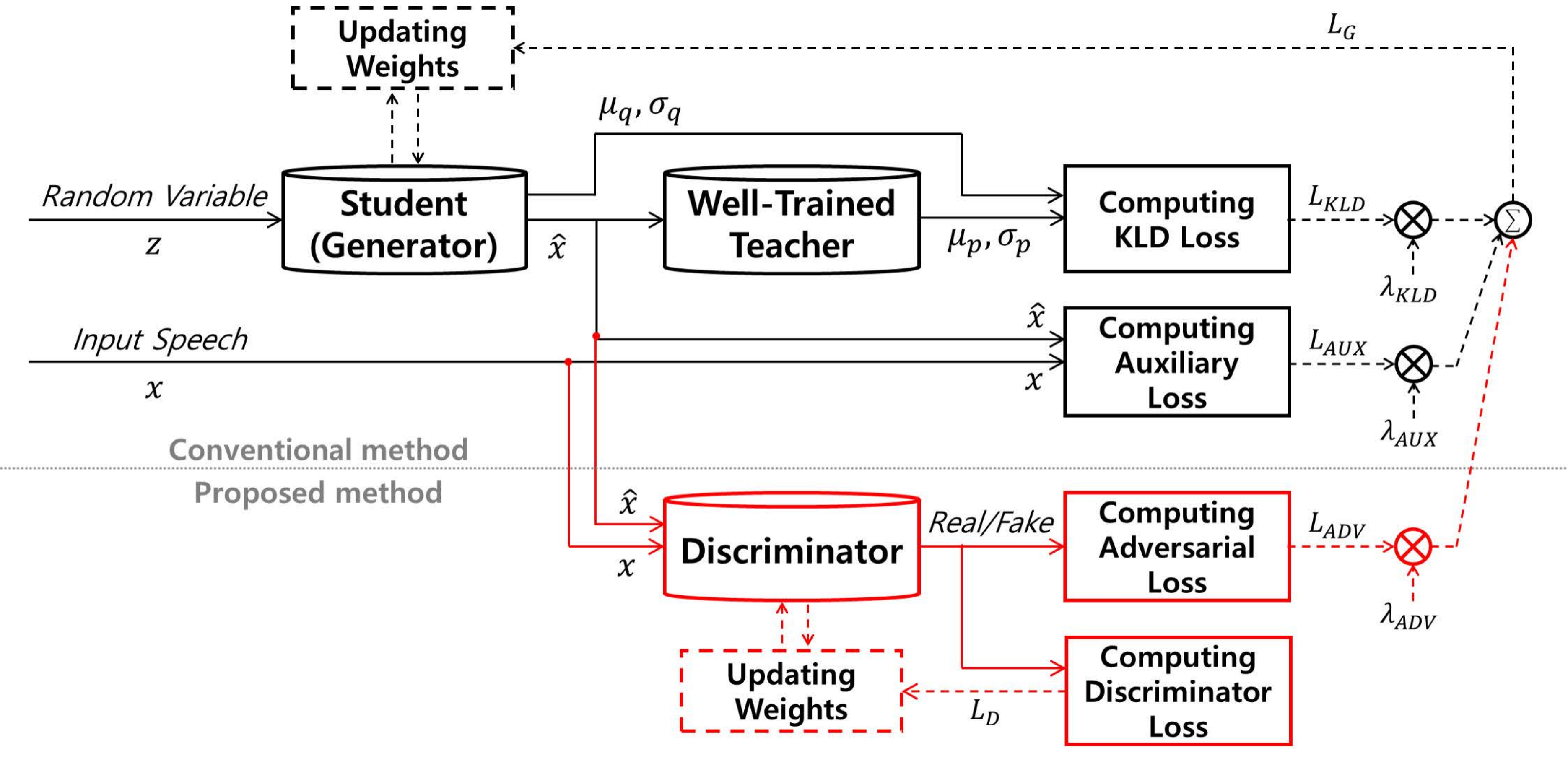,width=110mm}}
  \caption{An illustration of the distillation process for parallel waveform generation, where our proposed teacher-student framework adds an adversarial training process to the conventional methods (upper only).}
  \label{fig:block}
\end{figure*}

\section{Related work}
\label{related work}
    The idea of using PWG methods in the WaveNet framework is not new.
    By minimizing the KLD between output distributions of the teacher and student, \textit{parallel WaveNet} successfully achieves to distill the IAF student model from the teacher WaveNet model \cite{Oord2018ParallelWF}.
    By combining regularized KLD distillation with frame-level STFT loss, \textit{ClariNet} has proposed an effective and stable training criterion \cite{Ping2018ClariNetPW}. 
    As the STFT-based loss function is designed to guide the IAF student to learn the time-frequency characteristics of speech signals, its output quality has been further improved.
    
    Meanwhile, GANs have attracted a great deal of attention in the speech signal processing community thanks to their capabilities to learn the distribution of realistic speech signals via adversarial training.
    The performance of the speech synthesis systems has been also significantly improved by implanting the GAN structure to the acoustic models \cite{zhao2018wasserstein, yang2017statistical, saito2018statistical, lee2018acoustic}, the post-filters \cite{Kaneko_2017_Interspeech}, and the glottal excitations \cite{Bollepalli2017, juvela2018waveform}.
    
    Our aim is to incorporate the adversarial learning method into the teacher-student training process to achieve high-quality PWG of speech signals.
    Although a prior work in using GAN structure in the PWG application has been undertaken \cite{tian2018generative}, our research differs from this study:
    The GAN in the prior work was not used to distill the student model from the teacher WaveNet, but used to adapt an already-trained student model to a specific speaker (e.g., a speaker adaptation task).
    On the other hand, we focus on the effect of the adversarial learning method in training the student model itself.
    We propose a generalized optimization criterion by combining conventional KLD distillation with frame-level STFT loss and the proposed GAN-based adversarial loss.
    
    In addition to above, our experiments seek to verify the superior performance of the proposed method over conventional PWG systems. 
    Furthermore, thanks to the GAN's good capability to represent the nature of speech signals, the quality of the synthesized speech from the student model becomes more natural than even that from the teacher WaveNet.

\section{Probability density distillation}
\label{parallel waveform gen}

\subsection{KLD distillation}

    Conventional teacher-student framework-based systems employ the KLD-based PDD method to transfer the knowledge of a well-trained autoregressive teacher WaveNet to the target IAF student model \cite{Oord2018ParallelWF,Ping2018ClariNetPW}.
    As the simplified architecture of the student model enables sampling the speech signal in parallel, the generation speed becomes much faster than that of the autoregressive teacher.
    
    The upper part of Figure~\ref{fig:block} depicts a distillation process of the conventional teacher-student framework.
    During the training process, the student model first transforms the input random variable $\bz$ to a waveform sample $\bxhat$, and is evaluated by the corresponding well-trained teacher WaveNet.
    The entire network of the student model is then optimized to represent the teacher's distribution by minimizing the regularized KLD between the output distributions of the teacher and the student as follows \cite{Ping2018ClariNetPW}:
    \begin{equation}
    L_{\mathrm{KLD}}(q,p) =
    \E_{\bz,\bxhat}\left[\sum_{t=1}^{T}\mathrm{KL}^{\mathrm{reg}}(q(\hat x_{t}|z_{<t}) \parallel p(\hat x_{t} | \hat x_{<t}))\right], \label{loss kld}
    \end{equation}
    where $q(\bxhat) \sim N(\mu_{q}, \sigma_{q})$ and $p(\bxhat) \sim N(\mu_{p}, \sigma_{p})$ denote the output distributions of the student and teacher, respectively.

\subsection{STFT-based auxiliary loss}

    In addition to KLD minimization, it is well known that incorporating additional auxiliary losses using the ground truth dataset is advantageous to distill the student model well \cite{kim2016sequence}.
    Note that synthesized speech often contains undesirable artifacts (e.g., whispering voices) when the student IAF is trained with KLD loss alone \cite{Ping2018ClariNetPW}.
    
    To address the aforementioned problem, loss functions that are correlated with the perceptual audio quality should be used to train the student model \cite{arik2019fast}.
    In this paper, we adapt a frame-level auxiliary loss between the original and the generated speech samples as follows:
\begin{equation}
    L_{\mathrm{AUX}}(q) = \E_{\bx, \bxhat} \left[L_{\mathrm{SC}}(\bx,\bxhat) + \lambda_{\mathrm{MAG}} L_{\mathrm{MAG}}(\bx,\bxhat)\right], \label{auxloss}
\end{equation}
    where $\bx$ and $\bxhat$ denote the target and the estimated speech signal; $\lambda_{\mathrm{MAG}}$ denotes a weight coefficient to balance two losses, \textit{spectral convergence} ($L_{\mathrm{SC}}$) and \textit{log STFT magnitude} loss ($L_{\mathrm{MAG}}$), which is defined as follows \cite{arik2019fast}:
\begin{equation}
     L_{\mathrm{SC}}(\bx,\bxhat) = \frac{\parallel |\mathrm{STFT}(\bx)| - |\mathrm{STFT}(\bxhat)|\parallel_{F}}{\parallel|\mathrm{STFT}(\bx)|\parallel_{F}},
\end{equation}
\begin{equation}
     L_{\mathrm{MAG}}(\bx,\bxhat) = \parallel \mathrm{log}|\mathrm{STFT}(\bx)| - \mathrm{log}|\mathrm{STFT}(\bxhat)| \parallel_{1},
\end{equation}
    where $\parallel \cdot \parallel_{F}$ and $\parallel \cdot \parallel_{1}$ denote the Frobenius and $L_1$ norms, respectively; $|\mathrm{STFT}(\cdot)|$ denotes the STFT magnitudes. 
    Because the spectral convergence loss emphasizes spectral peaks and the log STFT magnitude loss accurately fits spectral valleys \cite{arik2019fast}, using a linear combination of both losses is helpful to effectively distill the student from the teacher WaveNet.

\section{Probability density distillation with generative adversarial networks}
\label{gan parallel gen}

    The KLD distillation combined with the STFT auxiliary loss has shown the feasibility to enhance the distillation efficiency. 
    To further improve the performance of the student model, we propose to incorporate GAN-based loss into the teacher-student framework.

    Figure~\ref{fig:block} shows the proposed distillation process. The student model is incorporated as a generator and jointly optimized by minimizing the adversarial loss ($L_{\mathrm{ADV}}$) along with the KLD loss ($L_{\mathrm{KLD}}$) and auxiliary loss ($L_{\mathrm{AUX}}$) as follows:
\begin{align}
    L_{\mathrm{G}}(q, p, D) &= \lambda_{\mathrm{kld}} L_{\mathrm{KLD}}(q,p) + \lambda_{\mathrm{aux}} L_{\mathrm{AUX}}(q) \notag \\ 
    &+ \lambda_{\mathrm{adv}}L_{\mathrm{ADV}}(q, D), \label{gloss}
\end{align}
    where $\lambda_{\mathrm{kld}}$, $\lambda_{\mathrm{aux}}$ and $\lambda_{\mathrm{adv}}$ denote the normalized weight coefficients for the KLD, STFT auxiliary, and adversarial losses, respectively\footnote{
    If the weight $\lambda_{\mathrm{kld}}$ is zero, the optimization criterion is equivalent to adversarial training methods \cite{juvela2018waveform,tian2018generative}. On the other hand, if the weight $\lambda_{\mathrm{adv}}$ is zero, it is equivalent to conventional PDD with the STFT auxiliary loss \cite{Ping2018ClariNetPW}.
    }.
    The adversarial loss, which represents how the student model learns the speech distribution from the discriminator, is defined as follows:
\begin{equation}
    L_{\mathrm{ADV}}(q, D) = \E_{\bxhat \sim q}\left[(1 - D(\bxhat))^2\right],
    \label{eq:adv}
\end{equation}
    where $D$ denotes the discriminator\footnote{
    This framework adopts a least-squares GANs thanks to its stability during the training process \cite{mao2017least, tian2018generative,Bollepalli2017,pascual2017segan}.}. 
    During the training process, the student model tries to deceive the discriminator into recognizing the generated samples as \textit{real} ($D(\bxhat) \to 1$). 
    On the other hand, the discriminator is trained to correctly classify the generated sample to \textit{fake} while classifying the ground truth to \textit{real} ($D(\bx) \to 1$) using the following optimization criterion:
\begin{equation}
    L_{\mathrm{D}}(q,D) = \E_{\bx \sim p_{\mathrm{data}}}[(1 - D(\bx))^2] + \E_{\bxhat \sim q}[D(\bxhat)^2], \label{dloss}
\end{equation}
    where $p_{\mathrm{data}}$ denotes the distribution of the speech signals.
    
    The entire training process encourages the student model to learn the distribution of the realistic speech waveform, which enables to generate more natural speech.
    Furthermore, the joint optimization with conventional distillations can address the limitations of feedforward GAN to capture the sample-level correlations of the speech signal \cite{juvela2018waveform}.
    Consequently, the perceptual quality of synthesized speech generated by the proposed method is effectively improved. 
    

\section{Experiments}
\label{experiments}

\subsection{Experimental setup}

    To investigate the effectiveness of the proposed method, we trained student models using the following four different optimization criteria:
\begin{itemize}
    \item {\bf AX}: Auxiliary loss.
    \item {\bf AXAD}: Auxiliary and adversarial losses.
    \item {\bf KLAX}: KLD and auxiliary losses.
    \item {\bf KLAXAD}: KLD, auxiliary and adversarial losses.
\end{itemize}

    In the experiments, we used a phonetically and prosaically balanced speech corpus recorded by a female professional Japanese speaker. 
    The speech signals were sampled at 24 kHz, and each sample was quantized by 16 bits. 
    In total, 3,299 utterances (7.34 hours) were used for training, 412 utterances (0.89 hours) were used for development, and another 413 utterances (0.92 hours) not included in either the training or development steps were used for evaluation.
    The leading and trailing silences in the speech signal were trimmed using a pre-processor, and 80-band log-mel spectrograms were extracted for composing the conditioning feature vectors.
    The frame and shift lengths were set to 25 ms and 5 ms, respectively. 
    Before training, the conditional feature vectors were normalized to have zero mean and unit variance.
    All the models and experiments were implemented using NAVER smart machine learning (NSML) platform \cite{kim2018nsml}.

    The teacher model was Gaussian autoregressive WaveNet \cite{Ping2018ClariNetPW}, which consisted of 24 layers of dilated residual convolution blocks with four exponentially increasing dilation cycles. 
    The number of residual channels, skip channels were 128 and convolution filter size was 3. 
    The conditioning features were upsampled by nearest neighbor upsampling followed by 2-D convolutions for interpolation \cite{odena2016deconvolution}. 
    The upsampling was split into five modules. The scales were [2, 2, 2, 3, 5]. 
    The kernel sizes for the 2-D convolutions were set to $2s + 1$, where $s$ denotes the upsampling scale.
    The teacher model was trained for 1 M steps with an \textit{Adam} optimizer \cite{kingma2014adam}.
    The initial learning rate was set to $0.001$, and it was reduced by half for every 200 K steps.
    The minibatch size was eight and the length of each audio clip was 12 K time samples.

\begin{table}[t]
  \caption{Normalized weight coefficients for training the different IAF student models.}
  \label{tab:weights}
  \vskip -3pt 
  \centering
  \begin{tabular}{ lccc }
    \toprule
    \multicolumn{1}{l}{\textbf{Method}} &
    \multicolumn{1}{c}{\textbf{$\lambda_{\mathrm{kld}}$}} &
    \multicolumn{1}{c}{\textbf{$\lambda_{\mathrm{aux}}$}} &
    \multicolumn{1}{c}{\textbf{$\lambda_{\mathrm{adv}}$}} \\
    \midrule
    AX       & - & $1.00$ & - \\
    AXAD    & - & $0.33$ & $0.67$       \\ 
    KLAX     & $0.09$ & $0.91$ & -      \\ 
    KLAXAD  & $0.03$ & $0.32$ & $0.65$     \\  
    \bottomrule
  \end{tabular}
  \vskip -10pt
\end{table}

    The student models were based on Gaussian IAFs \cite{Ping2018ClariNetPW}, each consisted of six flows in our settings. 
    Each flow was parameterized by a WaveNet that had ten layers of dilated residual convolution blocks with an exponentially increasing dilation cycle. 
    The number of residual channels, skip channels was 64 and filter size was 3. 
    The architecture of the upsampling network was the same as that of the teacher, and all the weights were initialized by the teacher's.
    The IAF student models were trained for 500 K steps with an Adam optimizer.
    The normalized weight coefficients (i.e., $\lambda_{\mathrm{kld}}$, $\lambda_{\mathrm{aux}}$ and $\lambda_{\mathrm{adv}}$) for training the different IAF student models are summarized in Table \ref{tab:weights}. 
    The initial learning rate was set to $0.0001$, and it was reduced by half for every 200 K steps.
    The minibatch size was eight and the length of each audio clip was 20.4 K time samples. 
    The STFT auxiliary loss was computed with a 25 ms Hanning window with 5 ms shift. The weight $\lambda_{\mathrm{MAG}}$ in Equation \ref{auxloss} was set to $1/(N_{\mathrm{frame}} \times F_{\mathrm{freq}})$, where $N_{\mathrm{frame}}$ and $F_{\mathrm{freq}}$ denote the number of time frames and frequency bins of the STFT magnitude, respectively.
    The KLD loss was computed as the same way as ClariNet \cite{Ping2018ClariNetPW}.

    In the proposed adversarial learning method, the discriminator consisted of ten layers of non-causal dilated 1-D convolutions interleaved with leaky ReLU activation function ($\alpha = 0.2$). The strides for the 1-D convolutions were set to 1 and linearly increasing dilations were applied for the 1-D convolutions\footnote{
    Our preliminary experiments verified that the linearly increasing dilations performed better than the exponentially increasing receptive fields for the discriminator.} starting from 1 to 8 except for the first and last layers.
    %
    Scalar predictions per-time step were done to better capture sample-level detailed differences between generated and real samples, and then averaged to compute the discriminator loss.
    The number of channels and filter size were 64 and 3, respectively. The conditioning feature vectors were not used for the discriminator.
    Because it is impossible to optimize the discriminator directly at the beginning of the training process, the student model as a generator was trained without the adversarial loss during the first 200 K steps.
    After warmup, the discriminator was sequentially optimized for 50 K with an Adam optimizer and finally entire networks were jointly trained via the adversarial learning method for the remaining 300 K steps. 
    The initial learning rate for the discriminator was set to $0.00005$, and it was reduced by half for every 200 K steps.

\subsection{Experiment results}


\begin{figure}[t]
  \vskip -7pt
  \centering
  \centerline{\epsfig{figure=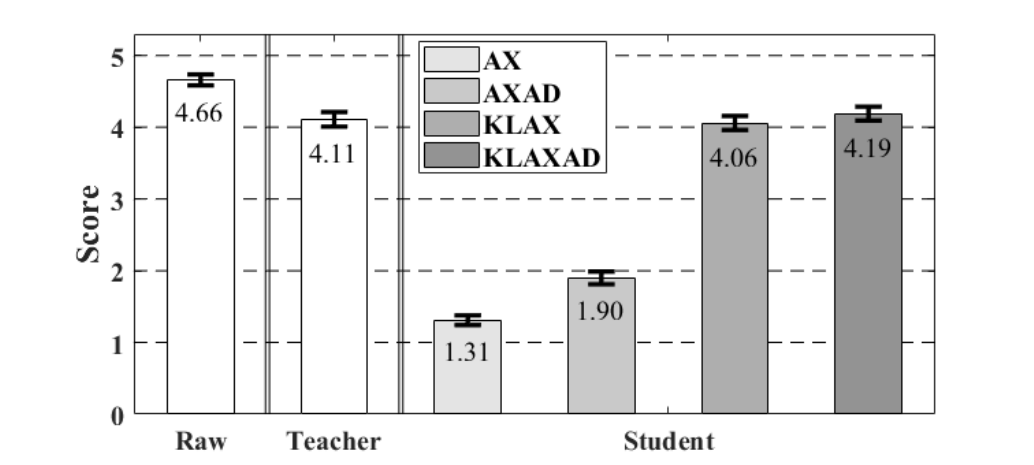,width=95mm}}
  \caption{The MOS results with 95 \% confidence intervals.}
  \label{fig:mos}
\end{figure}

    To evaluate the perceptual quality of the proposed system, mean opinion score (MOS)\footnote{Generated audio samples are available at the following URL:\\     \url{https://r9y9.github.io/demos/projects/interspeech2019/}} tests were performed.
    Fourteen native Japanese speakers were asked to make quality judgments about the synthesized speech samples using the following five possible responses: 1 = Bad; 2 = Poor; 3 = Fair; 4 = Good; and 5 = Excellent. 
    In total, 20 utterances were randomly selected from the test set and were then synthesized using the different generation models. 
    
    Figure~\ref{fig:mos} shows the MOS test results with respect to different generation models.
    The findings can be summarized as follows: (1) The IAF student trained only with the auxiliary loss (i.e., AX) performed worst. 
    Although adding the adversarial loss (i.e., AXAD) proved advantageous to improve the perceptual quality of the synthesized speech, it still scored poorly since it was challenging to capture sample-level correlations of the speech signal without the KLD-based distillation criterion.
    This can be confirmed by the test results for the KLAX system, where the perceptual quality was significantly improved by using the KLD distillation criterion with the auxiliary loss.
    (2) Among the IAF students, the proposed adversarial training method (i.e., KLAXAD) achieved the best quality. 
    In particular, the proposed system outperformed even the teacher WaveNet model.
    This was because the adversarial training guided the IAF student to learn the distribution of realistic speech waveform.
    Consequently, the proposed system with the adversarial training method achieved 4.186 MOS.

    To further verify the effect of the proposed method, we designed additional experiments by refining the normalized weight coefficients of the proposed system (i.e., $\lambda_{\mathrm{kld}}$, $\lambda_{\mathrm{aux}}$ and $\lambda_{\mathrm{adv}}$ in the KLAXAD system).
    Note that the previous listening test results verified that it is necessary to use the KLD-based distillation criterion during the training process. However, the best achievable quality of the distilled student model can be limited to the teacher model if the weight for KLD-based distillation (i.e., $\lambda_{\mathrm{kld}}$) is too large.
    Therefore, when the IAF student model starts to converge, it is recommended to decrease the $\lambda_{\mathrm{kld}}$ value.
    
\begin{figure}[t]
  \centering
  \includegraphics[width=\linewidth]{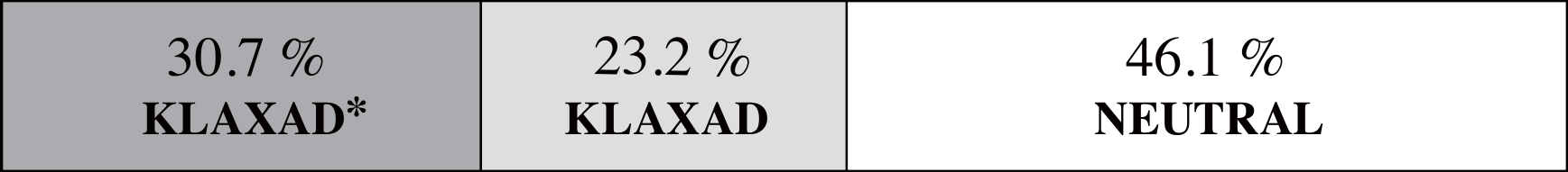}
  \caption{The test results of A/B/X preference comparison with two proposed methods including the baseline (KLAXAD) and its weight-refined version (KLAXAD$^{*}$).}
  \label{fig:abx}
\end{figure}

    Figure~\ref{fig:abx} depicts the A/B/X preference test results\footnote{The setups for the test were the same as for the MOS tests except that listeners were asked to rate the quality preference of the synthesized speech samples.
    }.
    Although the normalized weight coefficients were empirically modified ($\lambda_{\mathrm{kld}}$, $\lambda_{\mathrm{aux}}$, and $\lambda_{\mathrm{adv}}$ were set to zero, 0.33, and 0.67, respectively), the results confirm that weight-refined system (KLAXAD$^{*}$) provided better perceptual quality than the one originally proposed (KLAXAD).
    This implies that, when the student model started to converge, forcing the entire networks to be optimized toward the ground truth speech data rather than the teacher model was advantageous to generate more natural speech signal.
    Making the normalized weight coefficients learnable during the training process can further improve the general performance, which will be discussed in our future research.

\section{Conclusions}
\label{conclusions}

    This paper proposed an effective probability density distillation algorithm with generative adversarial networks (GANs) for WaveNet-based parallel waveform generation (PWG) systems. 
    Within a teacher-student framework, the proposed method incorporated an inverse autoregressive flow (IAF)-based student model as a generator in the GAN framework.
    Using novel optimization criteria based on adversarial learning method,
    the perceptual quality of the synthesized speech became much more natural.
    The experimental results verified that the PWG system using the proposed GAN-based training method performed better than the systems with conventional approaches.
    Despite the fact that the IAF student model was distilled from the teacher WaveNet, the merits of GAN to represent the nature of speech waveform enabled the student model to generate more natural speech than even the well-trained teacher model.

\section{Acknowledgements}
The work was supported by Clova Voice, NAVER Corp., Seongnam, Korea. The authors would like to thank Adrian Kim, Jaejun Yoo, Jung-Woo Ha, Lars Lowe Sj\"osund, Leonore Guillain and Xiaodong Gu at NAVER Corp., Seongnam, Korea, for their support.

\bibliographystyle{IEEEtran}
\bibliography{mybib}

\end{document}